\documentclass[instruments,article,accept,pdftex,moreauthors]{Definitions/mdpi} 

\firstpage{1} 
\pubvolume{6}
\issuenum{1}
\articlenumber{15}
\pubyear{2022}
\copyrightyear{2022}
\externaleditor{{Academic Editor: Antonio Ereditato} 
} 
\datereceived{20 January 2022} 
\dateaccepted{1 March 2022} 
\datepublished{6 March 2022} 
\hreflink{https://\linebreak \textls[-10]{doi.org/10.3390/instruments6010015}} 
\pdfoutput=1

\usepackage[detect-weight=true, detect-family=true]{siunitx}
\DeclareSIUnit{\sqrthz}{\ensuremath{{\text{\hertz}^{-1/2}}}}

\usepackage[justification=justified]{caption}

\usepackage{array}

\arrayrulecolor{black}

\Title{Characterization of Laser Systems at 1550 nm Wavelength for Future Gravitational Wave Detectors}

\TitleCitation{Characterization of Laser Systems at 1550 nm Wavelength for Future Gravitational Wave Detectors}

\Author{{Fabian Meylahn} 
 *\orcidA{} and Benno Willke \orcidB{}}

\AuthorNames{Fabian Meylahn and Benno Willke}

\AuthorCitation{Meylahn, F.; Willke, B.}

\address [1]{%
Max-Planck-{Institut} f\"ur Gravitationsphysik (Albert-Einstein-Institut) and {Institut} f\"ur Gravitationsphysik, \mbox{Leibniz Universit\"at Hannover}, Callinstr. 38, 30167 Hannover, Germany; {benno.willke@aei.mpg.de} 
}

\corres{\hspace{-0.75em}Correspondence: fabian.meylahn@aei.mpg.de}

 \abstract{The continuous improvement of current gravitational wave detectors (GWDs) and the 
 preparations for next generation GWDs place high demands on their stabilized laser sources. Some of the laser sources need to operate at laser wavelengths between \SI{1.5}{\um} and \SI{2.2}{\um} to support future detectors based on cooled silicon test masses for thermal noise reduction.
We present detailed characterizations of different commercial low power seed laser sources and power amplifiers at the wavelength of \SI{1550}{\nm} with respect to performance parameters needed in GWDs.
A combination with the most complete set of actuators was arranged as a master-oscillator power amplifier (MOPA), integrated into a stabilization environment and characterized. 
We present the results of this characterization that make this stabilized MOPA a highly relevant prototype for future GWDs as well as a low noise light source for other experiments in high precision metrology.}


\begin{document}

\section{Introduction}

Since the first direct detection of gravitational waves (GWs) from two merging black holes in 2015~\cite{Abbott2016S}, a multitude of further GW detections gave birth to the field of gravitational wave astronomy. One special signal arose from the merger of two neutron stars and was confirmed by multi-messenger astronomy~\cite{Abbott2017aS}. More than 70 GWs, documented in two GW detection catalogs~\cite{Abbott2019S,abbott2020S,Abbott2021}, were found in the last observation run of the ground-based gravitational wave detector (GWD) network~\cite{Buikema2020,Acernese2014,Affeldt2014}. 
The sensitivity of currently operating GWDs is fundamentally limited by quantum noise as well as coating thermal noise of the interferometer mirrors, often called test masses. Future GWDs need to reduce, in~particular, these noise contributions to achieve their anticipated ten fold sensitivity improvement.
The quantum noise at higher Fourier frequencies can be reduced by increasing  the laser power circulating inside the GWD interferometers  and can independently be further improved by the injection of squeezed light into the interferometer's output port~\cite{Tse2019, Acernese2019}.
The coating thermal noise is dependent on the coating properties and temperature.
By reducing the temperature of the test masses, the~coating thermal noise can generally be lowered, unless~an increase in the mechanical loss of the coatings at low temperature cancels the direct temperature effect~\cite{Harry2002}.
The first large scale GWD which uses cryogenic temperatures to lower the thermal noise is the KAGRA detector~\cite{Akutsu2019}, which uses sapphire test masses and a \SI{1064}{\nm} wavelength~laser.

Test masses made from silicon are likely to be used in future GWDs, as~they are available in high quality and larger sizes than sapphire.
As silicon is not transparent at the currently used wavelength of \SI{1064}{\nm}, this test mass material needs a laser wavelength between \SI{1.5}{\um} and \SI{2.2}{\um} to be successfully implemented in future GWDs~\cite{Adhikari2020}.
A European design study for a future GWD called the Einstein Telescope proposes three interferometers for the low frequency detection band and three for the high frequency detection band arranged in a triangular configuration. The~low frequency interferometers shall have silicon test masses and \SI{1550}{\nm} lasers, which can deliver \SI{3}{\watt} of optical power behind the input mode cleaner cavity~\cite{ETST2020}. This requirement translates to an output power of the lasers of roughly \SI{5}{\watt}.
The LIGO Voyager and Cosmic Explorer, future plans for GWDs in the US, shall use laser powers from \SI{200}{\watt} up to \SI{1}{\kilo\watt} with a laser wavelength compatible with silicon optics~\cite{Abbott2017b,LIGOSC2019}.

Ground-based GWDs generally place very high demands on their laser systems. The~laser sources have to be single frequency, single mode, linear polarization, and low noise. In~particular, the~frequency noise, power noise, and beam pointing noise in the detection band between a few hertz and typically a few kilohertz are most essential.
The second important frequency band is used for phase modulation of the interferometer's length and alignment control systems. This band spans from a few \si{\mega\hertz} up to the mid \si{\mega\hertz} range, depending on the chosen highest modulation frequency for the detector control. 
As no free-running laser meets the demanding GWD stability requirements, multiple stages of active feedback control and passive noise filtering are needed in the so-called laser pre-stabilization to prepare the lasers for injection into the GWD~interferometers.

Frequency stabilization of low power lasers in the range of \SI{1.5}{\um} wavelength were reported from various fields, such as the stabilization for a frequency comb~\cite{Dolgovskiy2013}, stabilization to ultra stable silicon cavities~\cite{Matei2017} or stabilization to a high Q fiber resonator~\cite{Bailly2020}.
A frequency and power stabilized system with fiber optics and an output power of \SI{10}{\milli\watt} was demonstrated~\cite{Takahashi2008}.
None of these experiments simultaneously achieves the GWD stability requirements for all relevant laser parameters and only parts of technologies needed for a GWD pre-stabilized laser system (PSL) are~demonstrated.

We used a well-tested laser characterization tool, the~so-called diagnostic breadboard (DBB), that was developed for \SI{1064}{\nm} \cite{Kwee2008} and adapted it for use at \SI{1550}{\nm}. With~this DBB, we characterized different seed lasers and power amplifiers. 
We first tested an external cavity diode laser from RIO~\cite{RIO} and compared our measurements with already published results on this seed laser~\cite{Numata2010,Tsuchida2011}. 
Fiber lasers from NKT Photonics~\cite{NKT} with different amplifiers and a seed source from NP Photonics~\cite{NPP} were characterized subsequently.
In addition, we tested two low power amplifiers, a~booster optical amplifier from \mbox{Thorlabs~\cite{Thorlabs}}, and a fiber amplifier built in a collaboration with the Laser Zentrum Hannover \mbox{e.V.~\cite{LZH}}.
Their free-running laser noise and their actuators for potential use in stabilization control loops are discussed in Section~\ref{sec:laserChar}.
In Sections~\ref{sec:setup_psl} and \ref{sec:laserStab}, we report on active feedback stabilized systems in order to demonstrate the usability of the optimal MOPA combination in a GWD~PSL.

With this work, we present, to~our knowledge, the~first stabilized laser system at \SI{1550}{\nm} wavelength with a simultaneous frequency and power stabilizations adequate for use in future~GWDs. 

\section{Experimental~Setup}
\label{sec:setup}

\begin{figure}[H]

\begin{adjustwidth}{-\extralength}{0cm}
\centering 
\includegraphics[width=\linewidth]{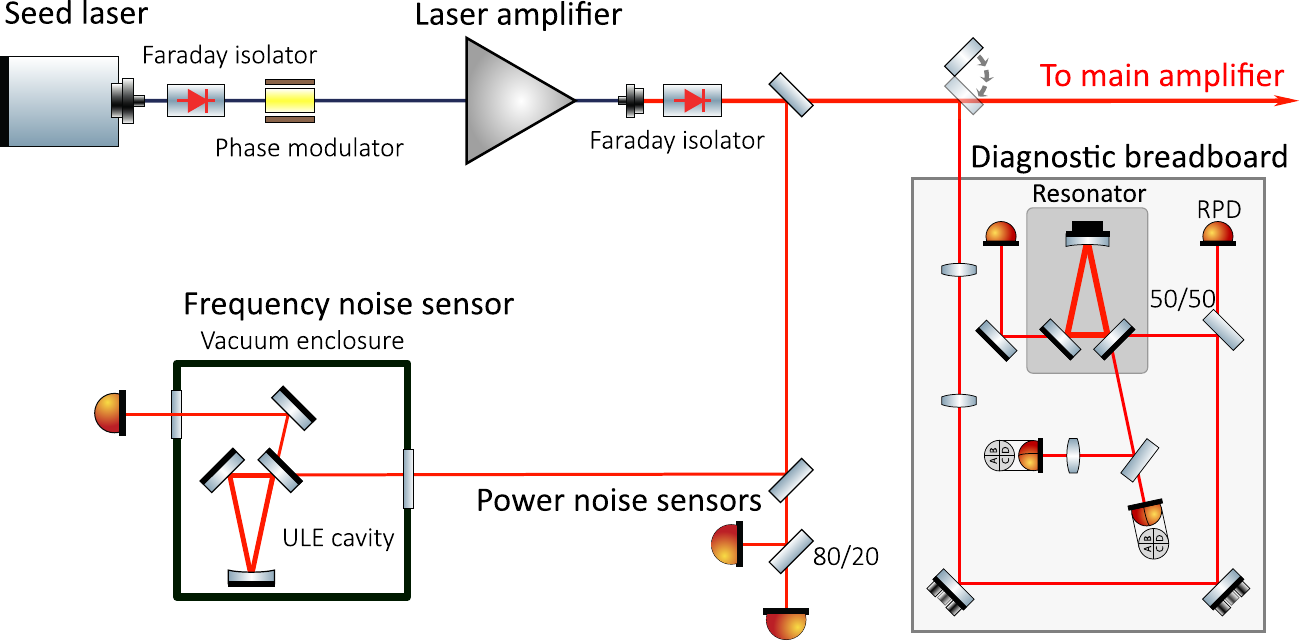}
\end{adjustwidth}
\caption{ The simplified stabilization setup shows the seed laser with \SI{1550}{\nm} wavelength, which was coupled via a fiber Faraday isolator to the fiber phase modulator and was amplified by the booster optical amplifier. A~Faraday isolator was placed in the \SI{115}{\mW} free space beam. This was to prevent reflections back to the amplifier. Parts of the beam were sent to power sensors and the frequency reference cavity. The~main fraction of the beam was analyzed with the diagnostic~breadboard.}
\label{fig:setup}
\end{figure}

For the laser characterization, a DBB for \SI{1550}{\nm} based on the original \SI{1064}{\nm} version~\cite{Kwee2008} was built. We used \SI{1550}{\nm} optics and germanium quadrant photodiodes and increased the transimpedance gains of the photodiodes. The~increased transimpedance gains were required to allow measurements at input powers of \SI{5}{\mW} to \SI{35}{\mW} as compared to the standard DBB with maximal \SI{150}{\mW} input power. This change is necessary to characterize the seed lasers without any power amplifier. A~schematic layout of the DBB can be seen in the right part of Figure~\ref{fig:setup}.

With this DBB, we analyzed free space laser beams in their noise and spatial mode properties. A~beam is coupled via two adjustable mode-matching lenses and two piezo alignment mirrors to a triangular optical ring resonator.
A low noise InGaAs photodiode (RPD) in front of the resonator is used to perform power noise measurements at Fourier frequencies between \SI{1}{\hertz} and \SI{100}{\mega\hertz}. A~photocurrent of up to \SI{18}{\milli\ampere} can be detected.
An analysis of the power transmitted by the resonator while scanning its length gives the mode decomposition of the mode-matched incoming beam.
The frequency noise of the laser beam can be calculated from DBB measurements by stabilizing the resonator length with the dither modulation techniques on a resonance for the laser frequency. The~dither modulation is imprinted at \SI{1}{\mega\hertz} on the resonator's piezoelectric element. The~error signal for the length control loop is generated by demodulation of the light power measured in reflection of the resonator.
The error signal and the control signal of the length feedback stabilization are captured to calculate the frequency noise between Fourier frequencies of \SI{1}{\hertz} to \SI{500}{\kilo\hertz}.
An equivalent technique in combination with the quadrant photodiodes is used for differential wavefront sensing based alignment control loops to stabilize the beam pointing~\cite{Morrison94,Anderson84}. This also enables the measurement of beam pointing noise.
In addition to the noise measurements, the~calibrated sensors of the DBB are used for transfer function measurements of the laser modulation~inputs.

All characterizations of the different laser systems are performed in the same setup. For~all measurements, the~beam path and the environmental conditions, such as airflow, devices in operation in the lab, or room light conditions, are kept constant to achieve comparable results. The~measurements were performed repeatedly over more than one day to confirm a stable performance of the tested lasers. Modulation inputs of the lasers are terminated or disabled during the noise~measurements.

\unskip

\subsection{Characterized Laser~Systems}
An overview of the characterized seed lasers and laser amplifiers is given in Table~\ref{tab:laser}.
We have chosen different commercial seed lasers, which are based on different technologies and have promising data sheet specifications for use in the field of ground-based gravitational wave detection. Single mode, single frequency operation with high stability in power and frequency within the gravitational wave detection band, as well as in the \si{\mega\hertz} range, were the most important selection criteria. In~addition, an~option for frequency control is needed to further stabilize the laser~frequency.

\begin{table}[H]
\caption{{{Tested laser sources and amplifier.} 
}}
\begin{tabular}{m{4.8cm}m{5.25cm}m{2.5cm}}
\toprule
{\textbf{Product Name} \boldmath{$^a$}} & {\textbf{Technology}} & {\textbf{Output Power}} \\ \midrule
\multicolumn{3}{l}{ {Seed laser sources} 
}\\ 
\midrule
\mbox{\textbf{{Orion}} laser source}, \mbox{RIO3135-3-34-5} \cite{RIO} & external cavity diode laser (ECDL) & \SI{10}{\mW} \\ 
\midrule
\mbox{Koheras \textbf{{Adjustik}} E15}, \mbox{K822-125-102 version 1} \mbox{K822-125-102 version 2} \cite{NKT} & erbium fiber lasers &\SI{40}{\milli\watt} \\
\midrule
\textbf{{NPP Seed laser}} of a high power laser module~\cite{NPP} & erbium fiber laser & \SI{28}{\milli\watt} $^b$ \\
\midrule
\multicolumn{3}{l}{  Pre-amplifier}\\ 
\midrule
\textls[-25]{Booster optical amplifier (\textbf{BOA})}, BOA1004P~\cite{Thorlabs} & \textls[-25]{semiconductor waveguide amplifier} & \SI{117}{\mW} $^b$\\ 
\midrule
\textbf{Fiber pre-amplifier} \cite{LZH} & erbium fiber amplifier & \SI{50}{\mW} $^b$ \\ 
\midrule
\multicolumn{3}{l}{  Power amplifier}\\ 
\midrule
\mbox{NKT \textbf{Boostik 2W}}, \mbox{K532-015-100} \cite{NKT} & erbium fiber amplifier & \SI{2}{\watt}\\
\midrule
\mbox{NKT \textbf{Boostik 10W} $^c$}, \mbox{K532-015-120} \cite{NKT} & erbium fiber amplifier & \SI{10}{\watt}\\
\midrule
\mbox{RIO \textbf{Grande} laser}, \mbox{RIO1175-9-34-5} \cite{RIO} & erbium fiber amplifier with integrated ECDL seed laser & \SI{2}{\watt}\\
\bottomrule
\end{tabular}
  \label{tab:laser}
  \pbox{\columnwidth}{\footnotesize{$^a$~The \textbf{bold} marked names are used in the text as a designation for the laser and amplifiers, $^b$~Measured output power,  $^c$~Two amplifiers of this kind were characterized.}} 
\end{table}

Fiber laser and semiconductor-based seed lasers were tested. Both achieve single frequency operation with a Bragg reflector for distributed feedback~\cite{Numata2010,NKT}.
To pre-amplify the seed power, two different types of amplifiers were investigated.
A BOA based on a fiber coupled InP/InGaAsP quantum well layer structure in a waveguide structure with a measured output power of \SI{117}{\mW} was tested~\cite{Thorlabs}.
The characterized fiber pre-amplifier is core pumped at \SI{976}{\nm} wavelength with a Bragg grating stabilized laser diode and was built in cooperation with the Laser Zentrum Hannover e.V.~\cite{LZH}.
For both amplifiers, an~in-house built, low noise current source was used with a fast modulation input (based on~\cite{Erickson2008,Troxel2011}).
The Boostik power amplifiers were tested with different seed lasers to test their influence on the actuation and noise characteristics. To~distinguish between the two different Boostik amplifiers, we name them by their slightly different output power levels of \SI{10.4}{\watt} and \SI{11}{\watt} later in the text.
The tested Grande amplifier has an integrated seed source, which is similar to the Orion~laser.

\subsection{Stabilized Laser~System}
\label{sec:setup_psl}

The usability of the actuators for laser stabilization needed for GWDs is demonstrated with the external cavity diode laser (ECDL) and the booster optical amplifier with an output power of \SI{115}{\mW}. The~Orion laser and the BOA were identified as the best MOPA combination for this test as they offer actuators for frequency and power stabilization with a good range and bandwidth. A~simplified setup is given in Figure~\ref{fig:setup}.
The laser frequency was stabilized to an ultra low expansion glass ceramic (ULE) spacer based, triangular reference cavity with a finesse of roughly 59,000 and a free spectral range of \SI{702}{\mega\hertz}. Two plane mirrors are optically contacted to the spacer while the curved mirror is glued to it. The~cavity is placed with Viton vibration isolation pads in a vacuum tank at a pressure of \SI{1}{\milli\bar} to reduce acoustical noise coupling.
The laser frequency was stabilized with the Pound{\textendash}Drever{\textendash}Hall (PDH) sensing scheme to the cavity~\cite{Drever1983,Black2001}. 
A fiber phase modulator was integrated to add phase modulation side-bands at \SI{44}{\mega\hertz} for the PDH sensing and as a fast actuator in the frequency stabilization feedback control~loop.

For the power stabilization, in- and out-of-loop photodiodes with transimpedance amplifiers were used. The~photodiodes were Perkin Elmer C30642. Both readout electronics are based on photodiode sensors used in the Advanced LIGO PSLs~\cite{Kwee2012}. The~in-loop detector has an improved bias circuit and an updated low noise operational amplifiers to increase the possible bandwidth of the stabilization~\cite{Meylahn2020}.

\section{Laser~Characterization}
\label{sec:laserChar}

First, the~different free-running laser systems and MOPA combinations were characterized with the DBB.
The transversal mode purity and the pointing noise of all tested lasers were measured. All show a low higher order mode content, as~all lasers have polarization maintaining, single mode fiber output~ports. 

\begin{figure}[H]
\includegraphics[width=0.95\linewidth]{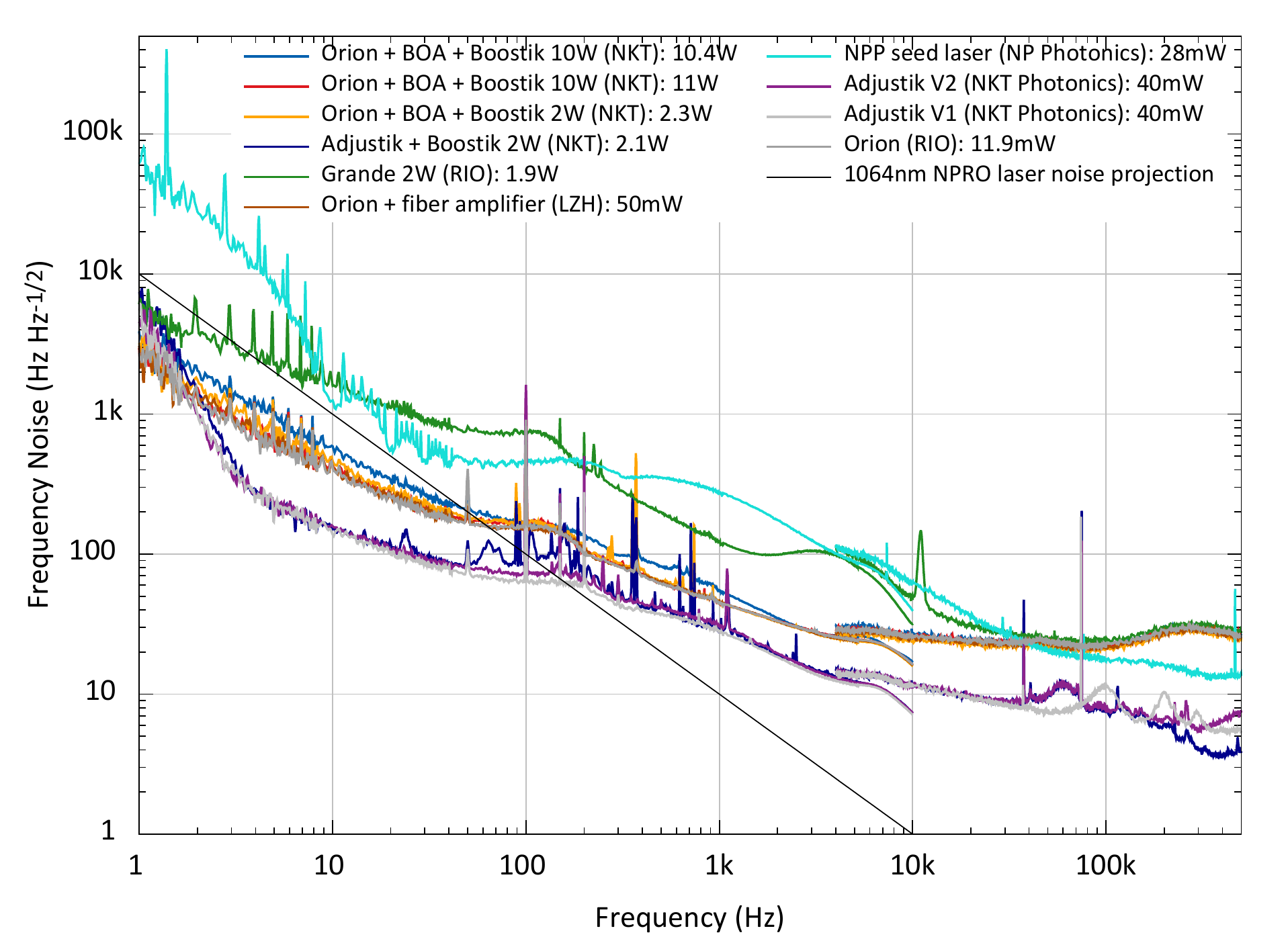}
\caption{ The free-running frequency noise of the different seed lasers and MOPA combinations is shown. The~frequency noise of the MOPA systems is dominated by their seed laser~noise.}
\label{fig:frq}
\end{figure}

The frequency noise measurements in Figure~\ref{fig:frq} show that only the seed lasers contribute in the measurement band to the total frequency noise of the modular systems. Small deviations at \SI{100}{\hertz} to \SI{300}{\hertz} are due to slightly different unity gain frequencies for the DBB resonator stabilization~loop.

All shown noise measurements are captured as a set of fast Fourier transformed measurements, which were averaged over multiple samples and normalized by the resolution bandwidth to achieve an amplitude spectral density. Multiple measurements with different spans were performed and stitched together to find the resolution for the logarithmic Fourier frequency scales.
The samples with different spans are plotted with a small overlap to account for transfer function uncertainties and different sensor whitening of different frequency~spans.

The lowest frequency noise is measured for the Adjustik lasers, followed by the Orion laser source. Both have a frequency noise comparable to the noise of the seed laser used in current GWD PSLs. The~seed lasers with \SI{1064}{\nm} wavelength used in current GWDs are non-planar ring oscillator lasers (NPRO). Their typical frequency noise is plotted as the NPRO projection~\cite{Kwee2008} in Figure~\ref{fig:frq}.

\begin{figure}[H]
\includegraphics[width=0.95\linewidth]{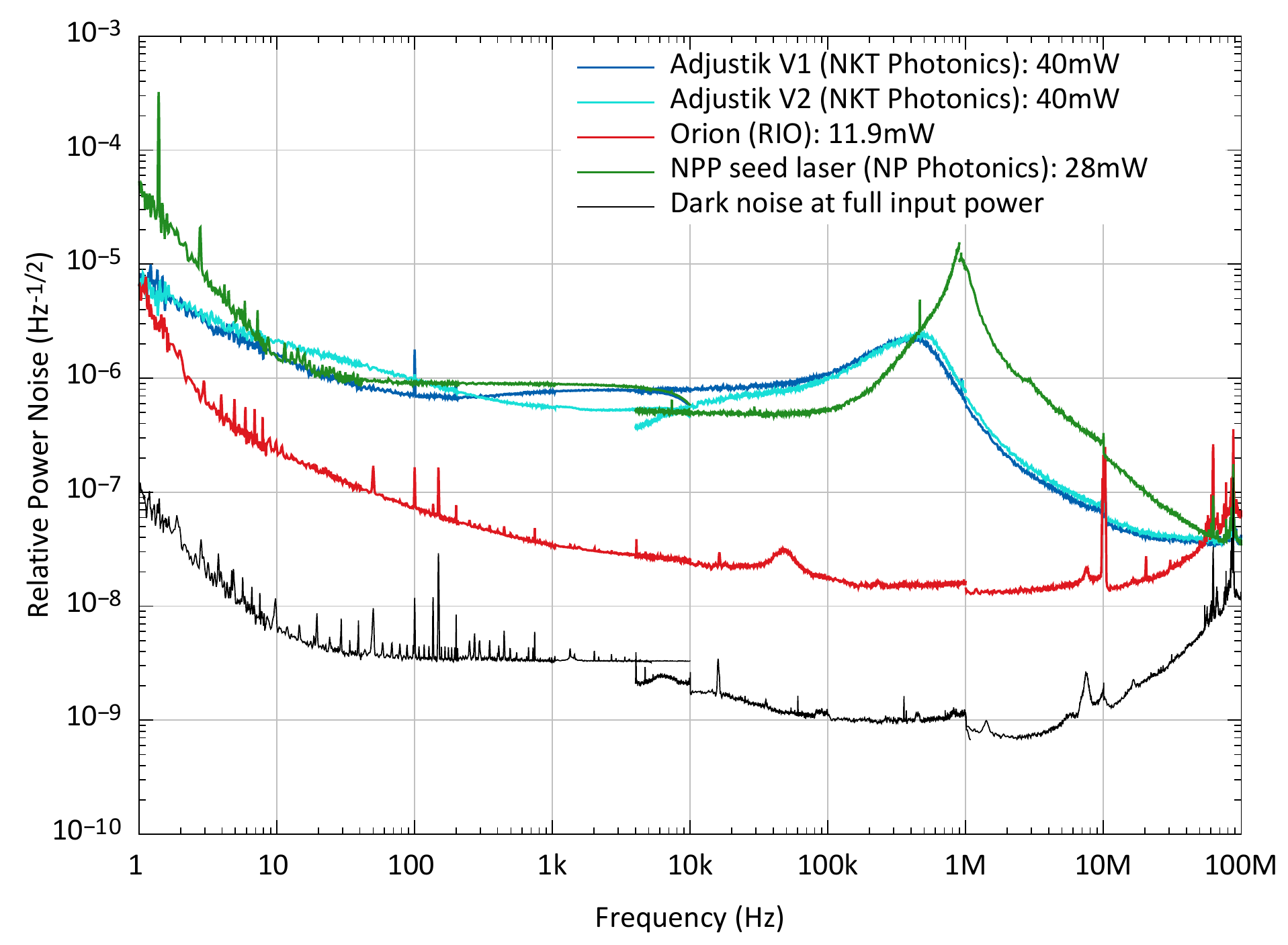}
\caption{ {The free-running} 
 RPN of the seed lasers, plotted as amplitude spectral density, follow two characteristic shapes. The~Orion laser has a very low free-running noise. The~fiber lasers have a higher noise floor and a characteristic relaxation oscillation~peak.}
\label{fig:rpnSeed}
\end{figure}

The relative power noise (RPN) of the seed lasers is shown in Figure~\ref{fig:rpnSeed} as an amplitude spectral density. 
The RPN of the Orion has a different spectral shape compared to fiber lasers. The~noise of the Orion laser roughly follows  a $f^{-1/2}$ slope down to the detection shot noise of \SI{1e-8}{\sqrthz}.  This low noise was achieved by replacing the supplied external \SI{5}{\volt} switching power supply by a linearly stabilized supply with lower noise coupling to the laser.
The RPN of the fiber lasers are, to our understanding, limited by the RPN of the used pump diodes, excess noise due to the re-absorption by the erbium ions, and the relaxation oscillation~\cite{Ralph1999,Cranch2003} within the laser resonator at \SI{500}{\kilo\hertz} for the Adjustik laser and at \SI{1}{\mega\hertz} for the NPP seed laser.
The peak at \SI{10}{\mega\hertz} in the measurements is due to a  timing signal for our data acquisition system. The~measurement of the Orion laser is influenced by dark noise above \SI{30}{\mega\hertz} due to low available laser~power.
\vspace{-3pt}

\begin{figure}[H]
  \includegraphics[width=0.9\linewidth]{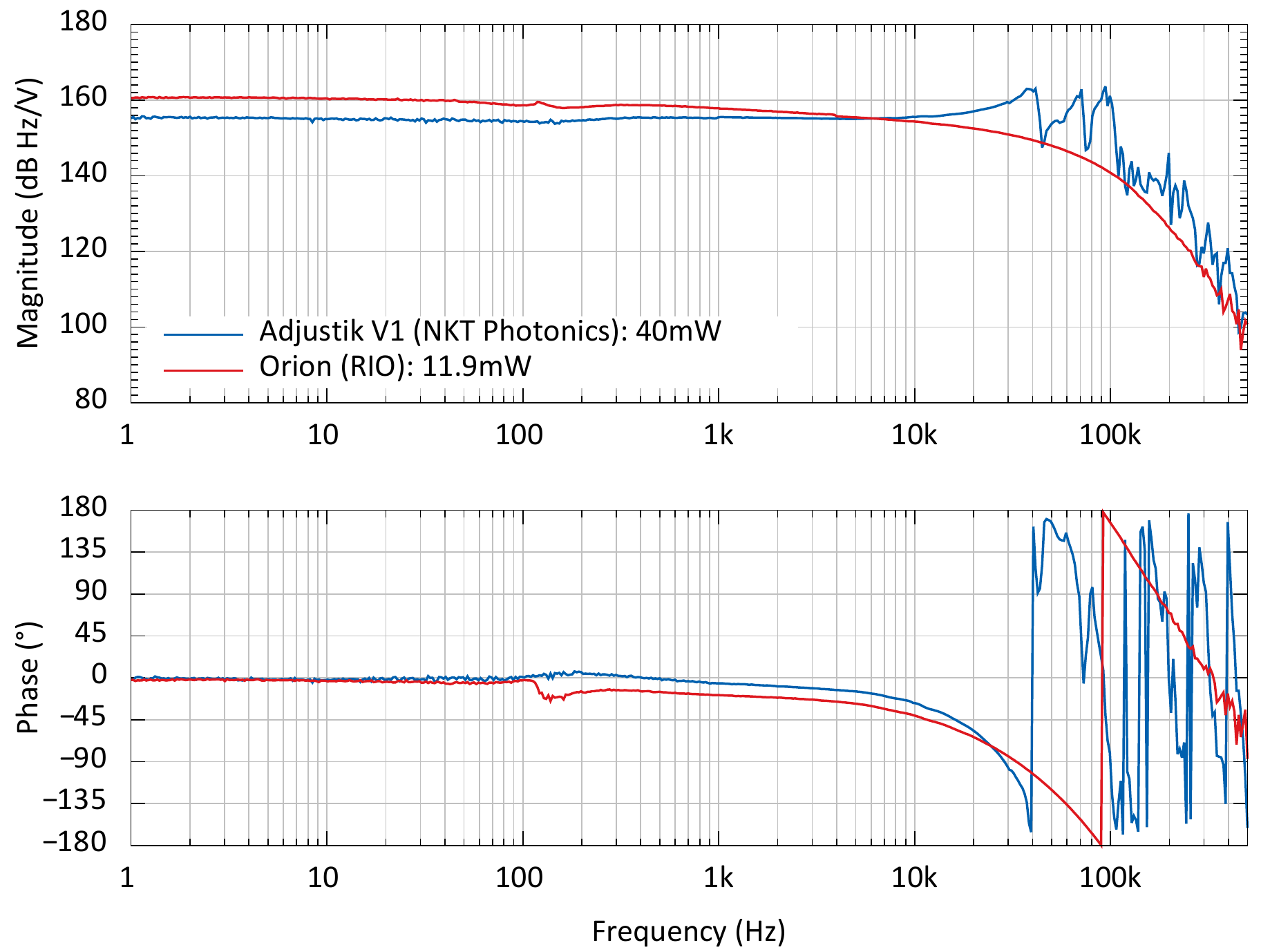}
  \caption{{The frequency} 
 actuation transfer functions of the fiber laser and of the Orion laser show a similar magnitude at low frequencies. Resonances in the fiber laser's transfer function reduce the usable actuation bandwidth significantly compared to the Orion~laser.}
  \label{fig:tfs_frqSeed}
\end{figure}

\begin{figure}[H]
  \includegraphics[width=0.90\linewidth]{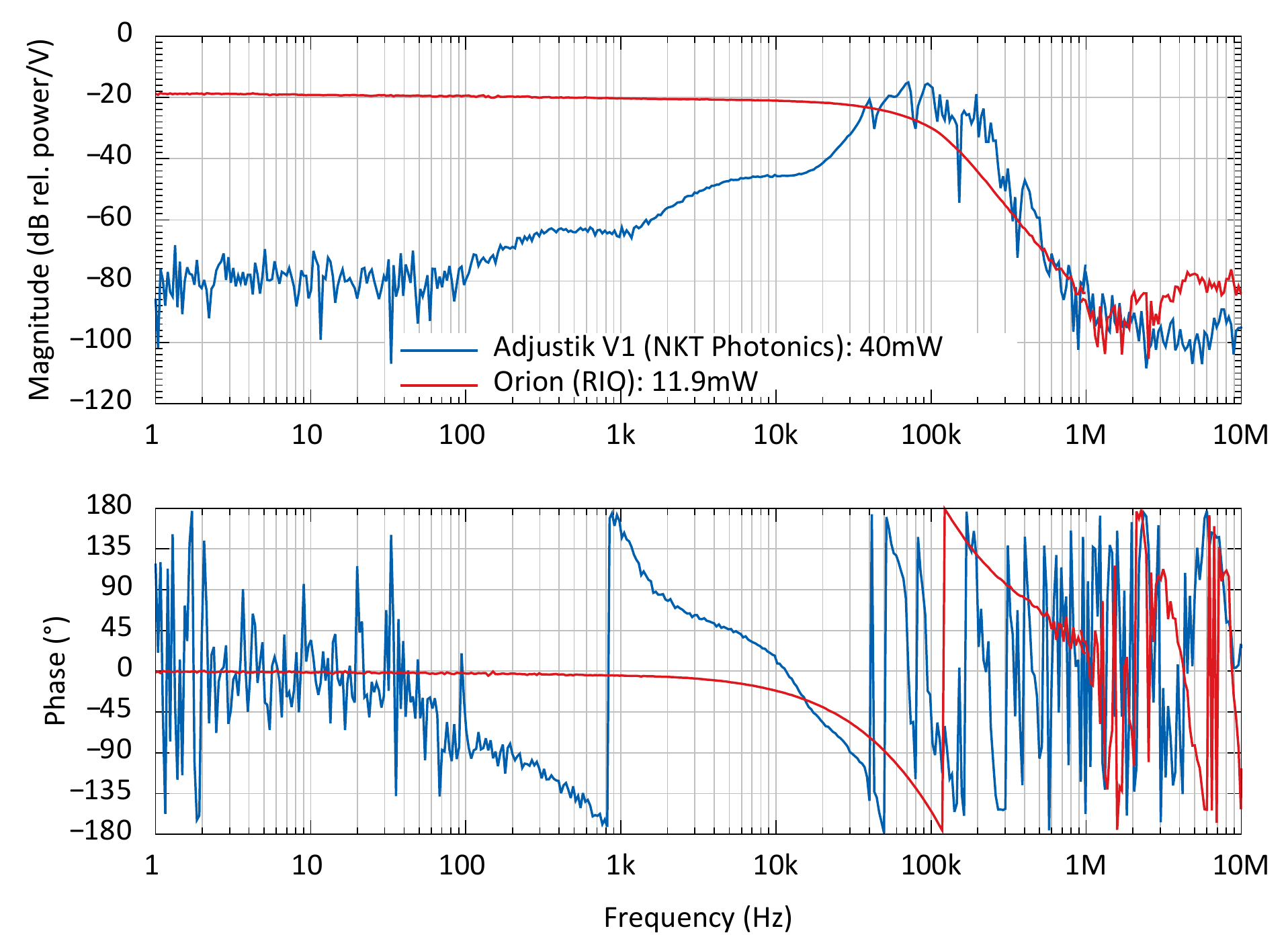}
  \caption{{The cross-couplings} 
 from frequency actuation to the output power of the Orion laser and fiber laser is shown. As~the pump current for the Orion laser also defines the output power, a~direct coupling is observed. Meanwhile, the~output power of the fiber laser shows a strong coupling only around \SI{30}{\kilo\hertz}.}
  \label{fig:tfs_power_crossSeed}
\end{figure}

The response and power cross-coupling of the frequency actuation inputs of the seed lasers are compared in Figures~\ref{fig:tfs_frqSeed} and~\ref{fig:tfs_power_crossSeed}. The~frequency of the fiber laser was tuned with a piezo electric element~\cite{NKTpzt}, which has characteristic resonances above \SI{40}{\kilo\hertz} and shows only a significant coupling from the frequency modulation to the output power of the laser at Fourier frequencies around \SI{70}{\kilo\hertz}. The~resonances limit the usable bandwidth of this actuator in feedback controls for a frequency~stabilization.

The frequency of the Orion laser is controlled by the pump current for the semiconductor, which introduces a carrier density dependent index of refraction change in the laser resonator. Hence, no mechanic resonances are present in the transfer function, but~a direct coupling to the optical gain and by that to the output power can be observed. The~transfer function decays with multiple poles, which can be easily compensated in a feedback control. This makes the delay equivalent phase loss in the transfer function to be the ultimate limit of the control bandwidth for a stabilization. 
A slow, high range tuning of the frequency is possible for both lasers by a temperature control of the laser~resonator.

\vspace{-6pt}

\unskip

\begin{figure}[H]
\includegraphics[width=0.95\linewidth]{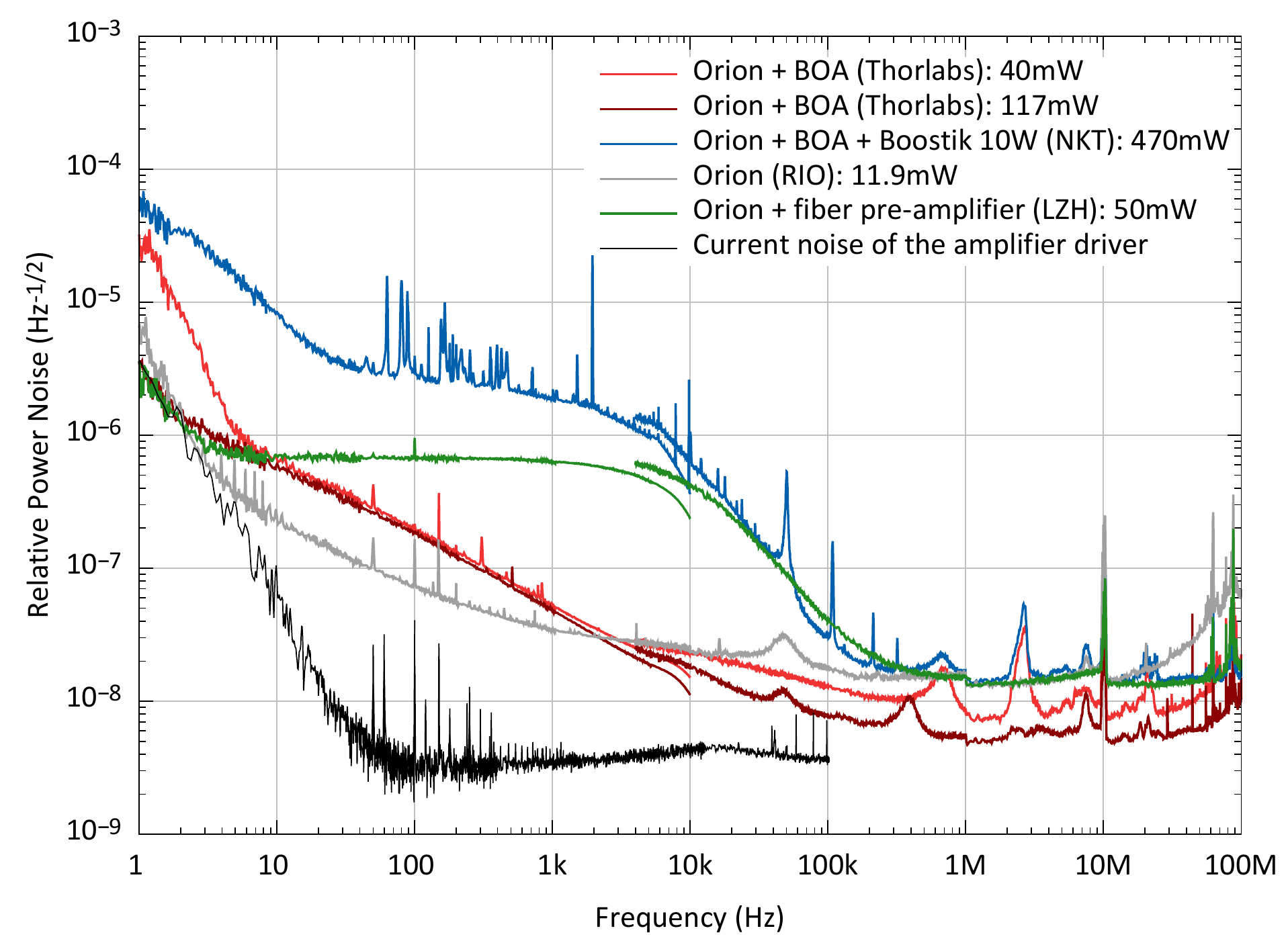}
\caption{ {The free-running} 
 RPN at the output of the pre-amplifiers is shown. For~all measurements, the~Orion laser is used as the seed source (grey curve).}
\label{fig:rpnLowAmp}
\end{figure}

\begin{figure}[H]
  \includegraphics[width=0.95\linewidth]{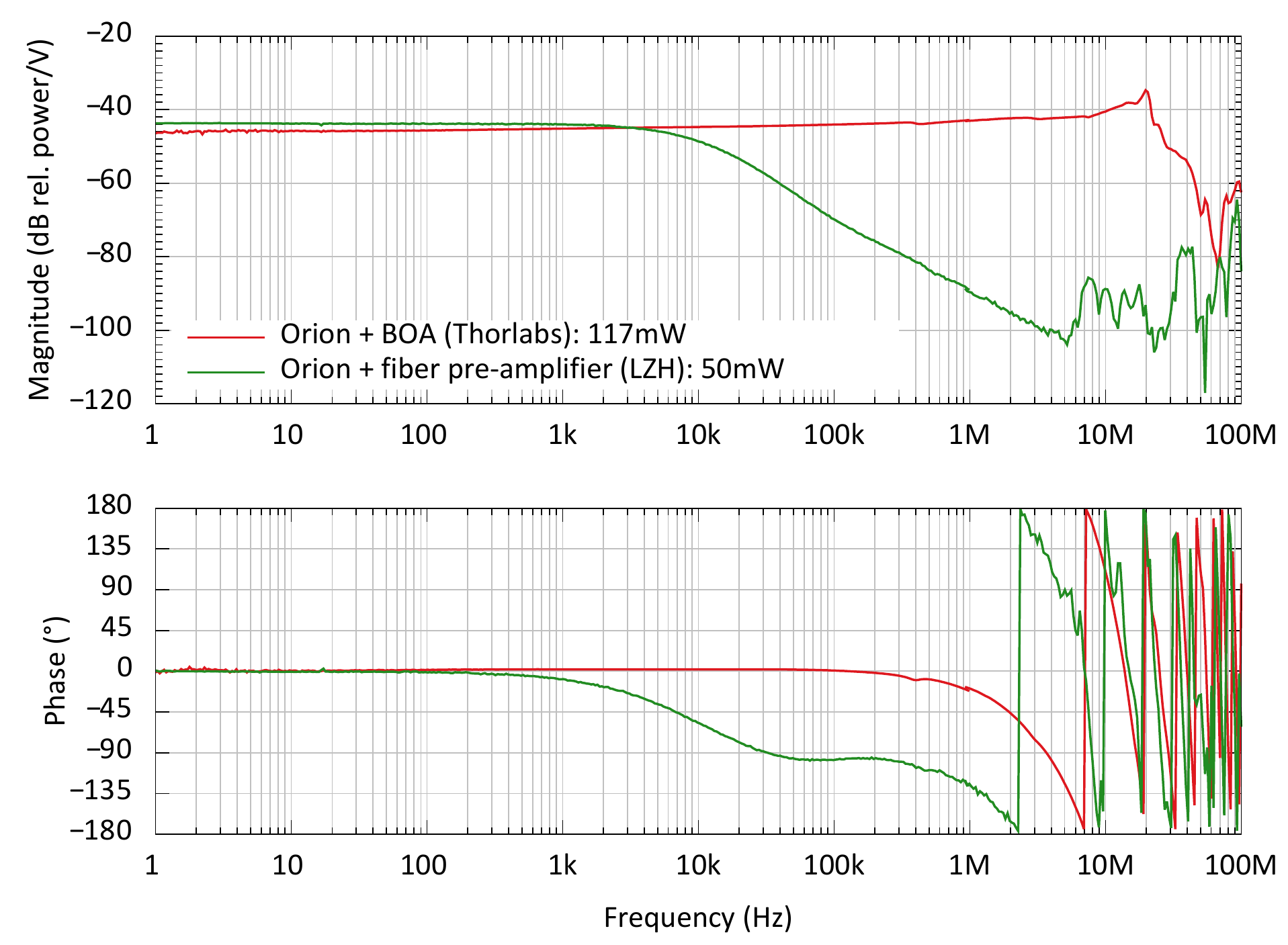}
  \caption{{The pump} 
 current of the BOA and fiber pre-amplifier's pump diodes is modulated to measure the transfer function to the pre-amplifier's output~power.}
  \label{fig:tfs_powerLowAmp}
\end{figure}

The RPN of the pre-amplifiers was tested with the Orion laser as seed laser, as~it has the lowest RPN. 
For the fiber pre-amplifier, as well as for the BOA, we used exactly the same low noise current driver with a relative current noise of \SI{3e-9}{\sqrthz} above \SI{100}{\hertz} (see Figure~\ref{fig:rpnLowAmp}). 
The higher RPN of the fiber pre-amplifier in comparison to the BOA is could be caused by reabsorption of the seed in the quasi three laser level system of erbium or amplified spontaneous emission \cite{digonnet2001} (pp. 68f., 543ff.). In~addition, we measured the RPN of the pump diodes for the fiber pre-amplifier to be at the same level as its output power noise. Tests with a pump diode, stabilized to more than a factor of 10 less RPN than the free-running pump light, were performed. 
These tests showed an improvement of less than a factor two in the output RPN of the amplifier.
Figure~\ref{fig:rpnLowAmp} shows the difference between BOA and fiber pre-amplifier systems. The~system based on the fiber pre-amplifier has a noise floor close to \SI{1e-6}{\sqrthz} for low frequencies.  At~a corner frequency of about \SI{10}{\kilo\hertz} the RPN decreases, which is due to the reduced response of the fiber amplifier to power noise of its pump source (see Figure~\ref{fig:tfs_powerLowAmp}).
In comparison, the~noise of the BOA pre-amplifier decreases roughly with a $f^{-1/2}$ slope. Measurements at the maximal output power of \SI{117}{\mW} and at \SI{40}{\mW}, which is the pre-amplifier's output power used in this work to seed the subsequent power amplifier, were taken with the BOA.
The relative current noise of the current supply for these two amplifiers is plotted in Figure~\ref{fig:rpnLowAmp} as the black curve. The~current noise is clearly below the power noise of both amplifiers and thereby does not limit their noise performance, assuming a linear coupling~mechanism.

The Boostik \SI{10}{\watt} amplifier from NKT consists of a pre-amplifier and a power amplifier. The~pre-amplifier was tested with the unpumped power amplifier.
Its noise (blue curve in Figure~\ref{fig:rpnLowAmp}) shows several sharp peaks that might be caused by internal electronics. The~noise level could also be increased due to the transmission through the unpumped active fiber, where reabsorption could happen.
The shot noise limits for the amplifier measurements shown in Figure~\ref{fig:rpnLowAmp} were between \SI{4e-9}{\sqrthz} and \SI{6e-9}{\sqrthz} and in the case of the Orion seed laser (grey curve) at \SI{1e-8}{\sqrthz}, as~only a lower optical power was available on the power sensor in the~DBB.

In Figure~\ref{fig:tfs_powerLowAmp}, the transfer functions of the pre-amplifiers from pump current to output power are shown. The~pump currents are modulated with \SI{5}{\milli\ampere\per\volt} for a total pump currents of \SI{522}{\milli\ampere} for the fiber amplifier and \SI{600}{\milli\ampere} for the BOA. The~transfer functions are calibrated as relative power modulation per input voltage. This transfer function typically forms a low-pass, where the corner frequency $f_\mathrm{eff}$ defines how fast the population of the upper laser level can adapt to the change in pump power~\cite{Tunnermann12}.
The corner frequency of the fiber pre-amplifier is determined by a fit to $f_\mathrm{eff}=$ \SI{6.7}{\kilo\hertz}. In~contrast, the~BOA's transfer function is limited by the speed of its current driver. The~typical corner frequency $f_\mathrm{eff}$ of these semiconductor-based  laser amplifiers is above \SI{100}{\mega\hertz} \cite{Ebeling1993} (p. 469).

\vspace{-4pt}

\begin{figure}[H]
  \includegraphics[width=0.95\linewidth]{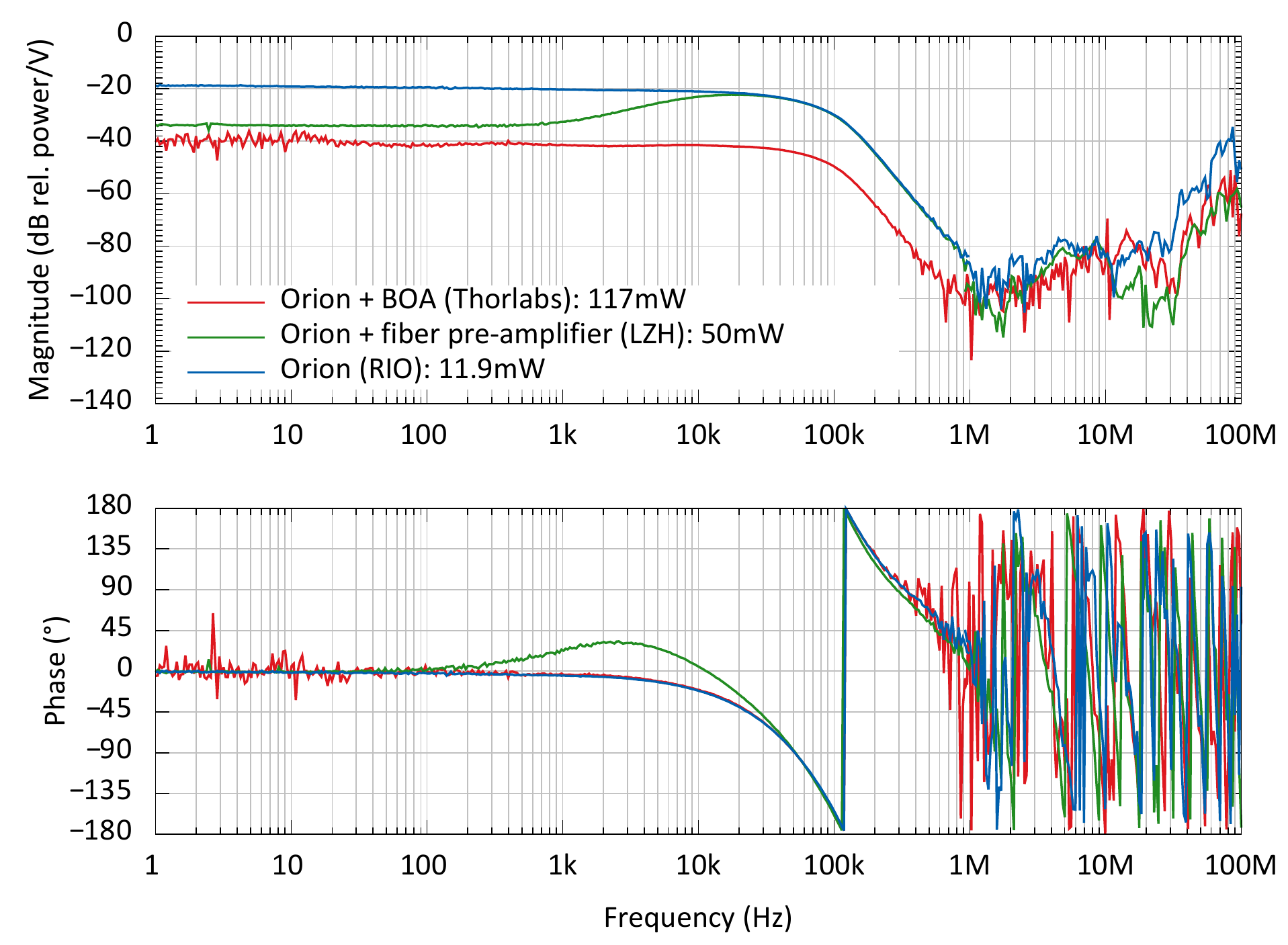}
  \caption{{The seed} 
 power is modulated to measure the transfer function to the output power of the pre-amplifiers. The~seed power modulation (blue curve) is transferred to the output of the pre-amplifiers (green and red curves).}
  \label{fig:tfs_power_crossLowAmp}
\end{figure}

The response of the fiber pre-amplifier to seed power modulation (see Figure~\ref{fig:tfs_power_crossLowAmp}) is suppressed at low frequencies with the amplifier gain, due to the seed power saturation of the amplifiers. Above~the corner frequency $f_\mathrm{eff}$, the~seed power modulation is amplified by the full power gain of the amplifiers~\cite{Tunnermann12}. The~seed power modulation sent to the BOA is suppressed in the whole bandwidth of the seed power modulator, due to the high corner frequency corresponding to this~amplifier.

\begin{figure}[H]
\includegraphics[width=0.95\linewidth]{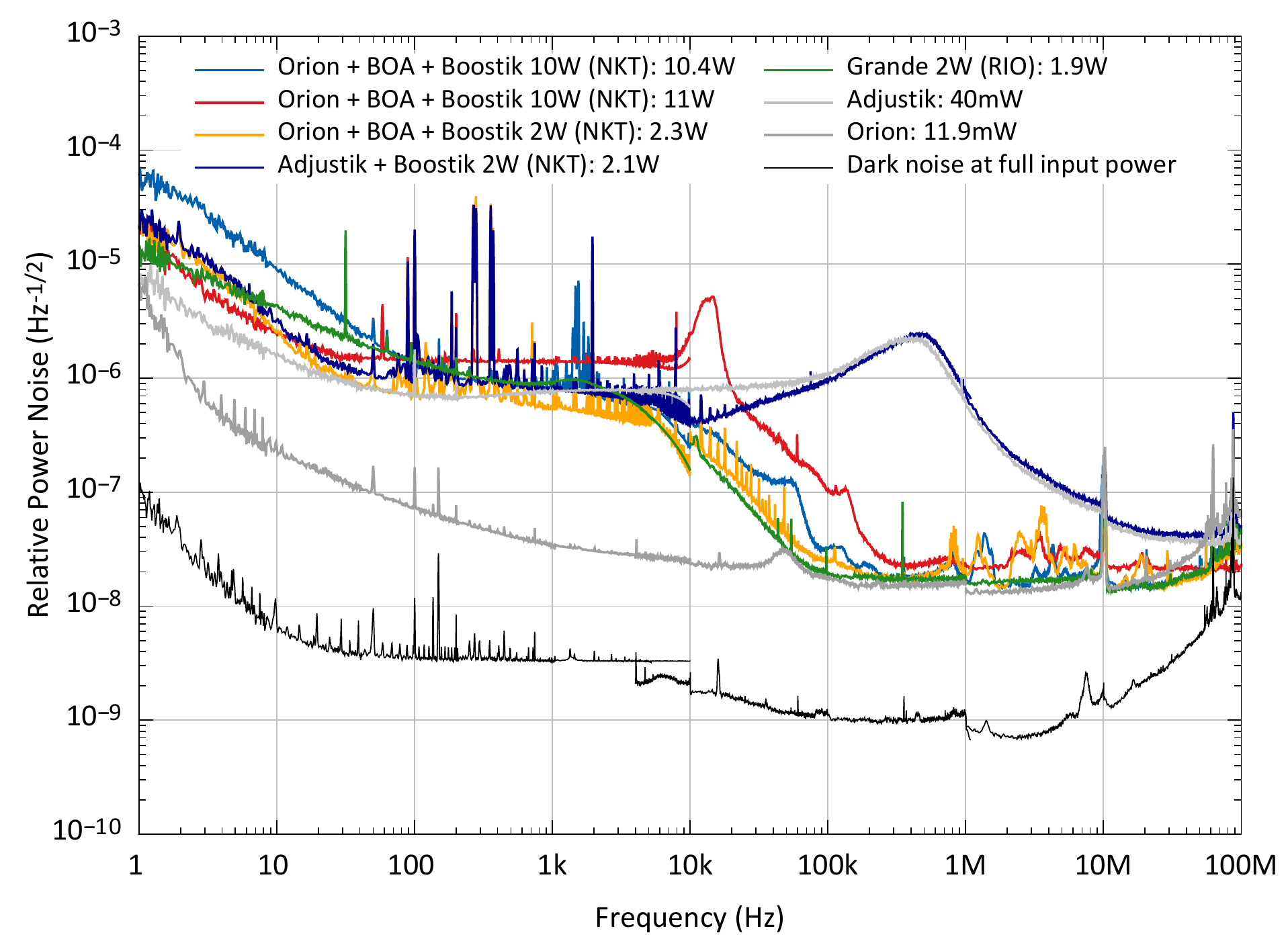}
\caption{ {The free-running} 
 RPN spectral densities at the output of the power amplifiers is shown. At~low frequencies, the~power noise of the power amplifiers is dominated by their pump power~noise. }
\label{fig:rpnHighAmp}
\end{figure}

The RPN of the all-fiber power amplifiers is plotted in Figure~\ref{fig:rpnHighAmp}.
The typical flat noise for frequencies up to \SI{10}{\kilo\hertz}, which was discussed for the pre-amplifiers, is again clearly visible. Above~\SI{10}{\kilo\hertz} the power noise differs with the used seed laser source.
We measured slightly different RPN levels for the two nominal similar \SI{10}{\watt} power level amplifiers, which might be caused by different noise levels of internal current drivers for the pump diodes.
The narrow peaks in the RPN measurements of the amplifier can be caused by internal electronics in the amplifier modules, such as temperature controllers or digital~controls.

\begin{figure}[H]
\includegraphics[width=0.90\linewidth]{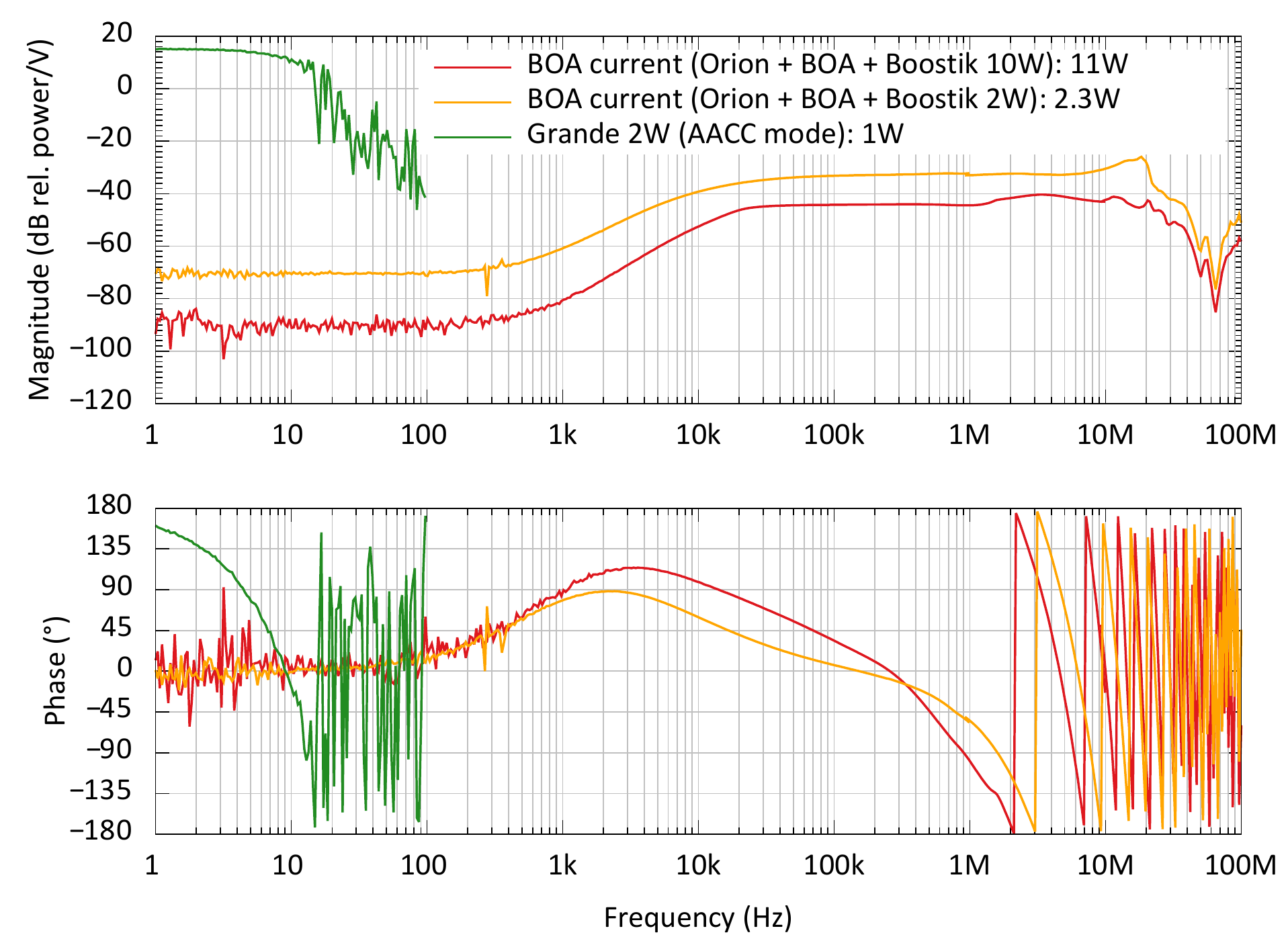}
\caption{{The transfer} 
 functions from seed power modulations to power modulations of the power amplifier output have the same shapes as the measurements with the fiber pre-amplifiers in Figure~\ref{fig:tfs_power_crossLowAmp}. }
\label{fig:tfs_powerHighAmp}
\end{figure}

The transfer functions in Figure~\ref{fig:tfs_powerHighAmp} show how the output power of the power amplifiers can be controlled by controlling their seed laser power.
The Grande MOPA laser does not allow us to change the power of its integrated seed separately from its frequency. We show the response of its analog power control for comparison. This input allows us to change the pump current with a bandwidth of less than \SI{10}{\hertz}. 
The other power amplifiers only provide  a slow digital control of their output power. Fast output power control is only possible via control of the seed or pre-amplifier power.
Due to the seed power saturation of the power amplifiers, they show a reduced response to power modulations of the seed below their corner frequency $f_\mathrm{eff}$ of the coupling transfer function. This makes it difficult to use only the pre-amplifier seed beam to stabilize the output power of such an~amplifier.

\unskip

\vspace{-3pt}

The relaxation oscillation frequency of the fiber seed laser is at or above the corner frequency of this transfer function. Thereby, its magnitude in a relative power noise spectral density is not attenuated, as~can be seen in the power amplifier's output noise in Figure~\ref{fig:rpnHighAmp}. 


The frequency modulation transfer functions from the pre-amplifier to the power amplifiers are up to small delay-driven phase losses given by the seed transfer function (see Figure~\ref{fig:tfs_frqSeed}). The~delay is caused by the propagation time through the optical fibers of the~amplifiers.

\section{Stabilization}
\label{sec:laserStab}

To demonstrate the capabilities of the actuators and sensors for a stabilized laser system, the~Orion laser with the BOA was stabilized. This combination was chosen due to its low noise and the available actuation capabilities. 
We did not include any power amplifier in our stabilization setup as they are not expected to influence the frequency noise and as the currently available models do not provide appropriate power~actuators. 

\begin{figure}[H]
\includegraphics[width=0.95\linewidth]{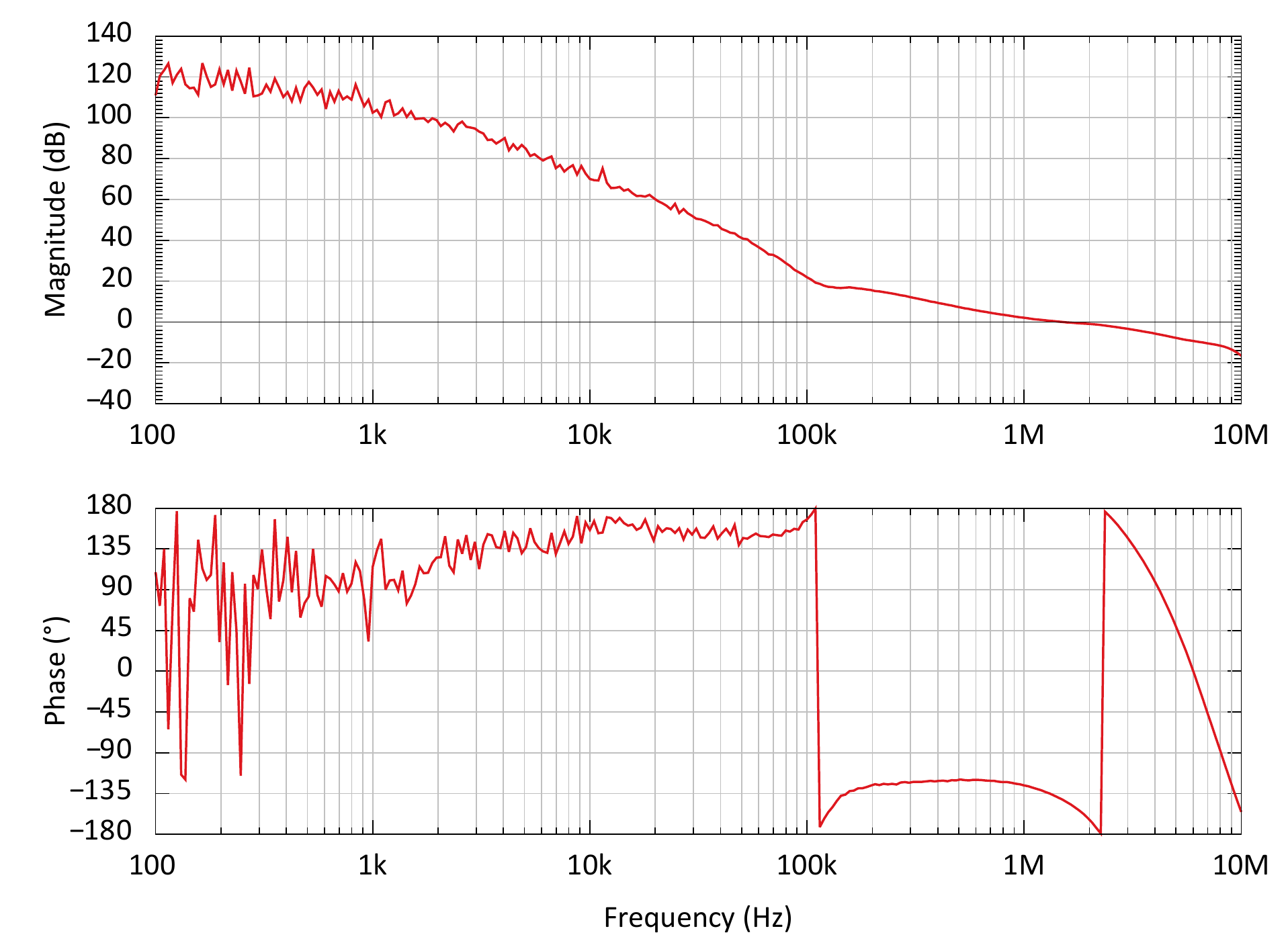}
\caption{{The open} 
 loop transfer function of the frequency stabilization of the laser to the resonance frequency of the reference cavity is shown. A~unity gain bandwidth of \SI{1.3}{\mega\hertz} was achieved by utilizing a cross-over between feedback to the seed laser and to the fiber phase modulator at \SI{100}{\kilo\hertz}.}
\label{fig:openloop_frq}
\end{figure}

The frequency stabilization is realized by acting on the ECDL's pump current and on the fiber phase actuator, which results in the open loop transfer function in Figure~\ref{fig:openloop_frq}. The~open loop transfer function was calculated from a disturbance reduction transfer function measurement, which describes  by how much an external noise source is reduced in the closed~loop.

Below \SI{100}{\kilo\hertz}, the pump current of the seed laser was used to control the laser frequency to be on resonance with the reference cavity.
At frequencies above \SI{100}{\kilo\hertz}, the fiber phase modulator was the dominating actuator for the stabilization. With~this combination, we achieved a unity gain bandwidth of \SI{1.3}{\mega\hertz}.
An applied voltage to the electro-optical phase actuator leads to a frequency actuation of $\SI{0.5}{\hertz\per\volt}\cdot f/\SI{1}{\hertz}$, which is a factor 100 higher than a typical free space phase modulator. In~comparison the frequency actuation of the laser has a gain of $\SI{100}{\mega\hertz\per\volt}$ (see Figure~\ref{fig:tfs_frqSeed}). This is a suitable actuator at low frequencies and the use of the fiber phase modulator is beneficial at frequencies above \SI{10}{\kilo\hertz}.
The measured open-loop gain shown in Figure~\ref{fig:openloop_frq} is limited at low frequencies by the dynamic range of the network analyzer. A~nested loop stabilization including an additional boost stage would allow for even higher loop gains below \SI{1}{\kilo\hertz}.
The bandwidth of this stabilization is limited by the required low-pass in the demodulation electronics of the PDH sensing and the electronic~delays.

\vspace{-6pt}

The control signal and the error signal are captured to compute the amplitude spectral density of the frequency noise. The~calibration of the control signals was derived by the actuator transfer functions measured with the DBB as a frequency sensor. The~error signal calibration was derived from the plant transfer function of the frequency stabilization, calculated from the measured noise suppression transfer function and the transfer function of the feedback controller. 
For the stabilized frequency noise measurements (Figure \ref{fig:stab_frq}), the~power stabilization was in operation (see below). No cross-couplings from the operating power stabilization to frequency noise were~found.

The control signal sent to the laser frequency actuator and to the phase actuator with the closed frequency control loop is consistent with the free-running frequency noise measured with the DBB. Only excess noise peaks at \SI{1.95}{\hertz} and harmonics were present due to ground loop coupling to the modulator input of the Orion laser. In~addition, a~ servo bump at \SI{2}{\mega\hertz} is visible.
The ground loop coupling is totally suppressed by the high loop gain and is not visible in the error signal noise. At~frequencies above \SI{100}{\kilo\hertz}, where the phase actuator is used for frequency control, an~electronic cross-coupling of the control signal to the error signal is visible (see the similarity of light blue and red curve in Figure~\ref{fig:stab_frq}). This is due to a common printed circuit board for error signal demodulation and fiber phase modulator~control.

The frequency noise of the stabilized system was also measured with the DBB and is shown in Figure~\ref{fig:stab_frq}.
This measurement was for frequencies below \SI{1}{\kilo\hertz} most likely limited by the length noise of the low finesse, piezo driven aluminum spacer cavity that serves as a frequency reference in the~DBB.

\begin{figure}[H]
\includegraphics[width=0.9\linewidth]{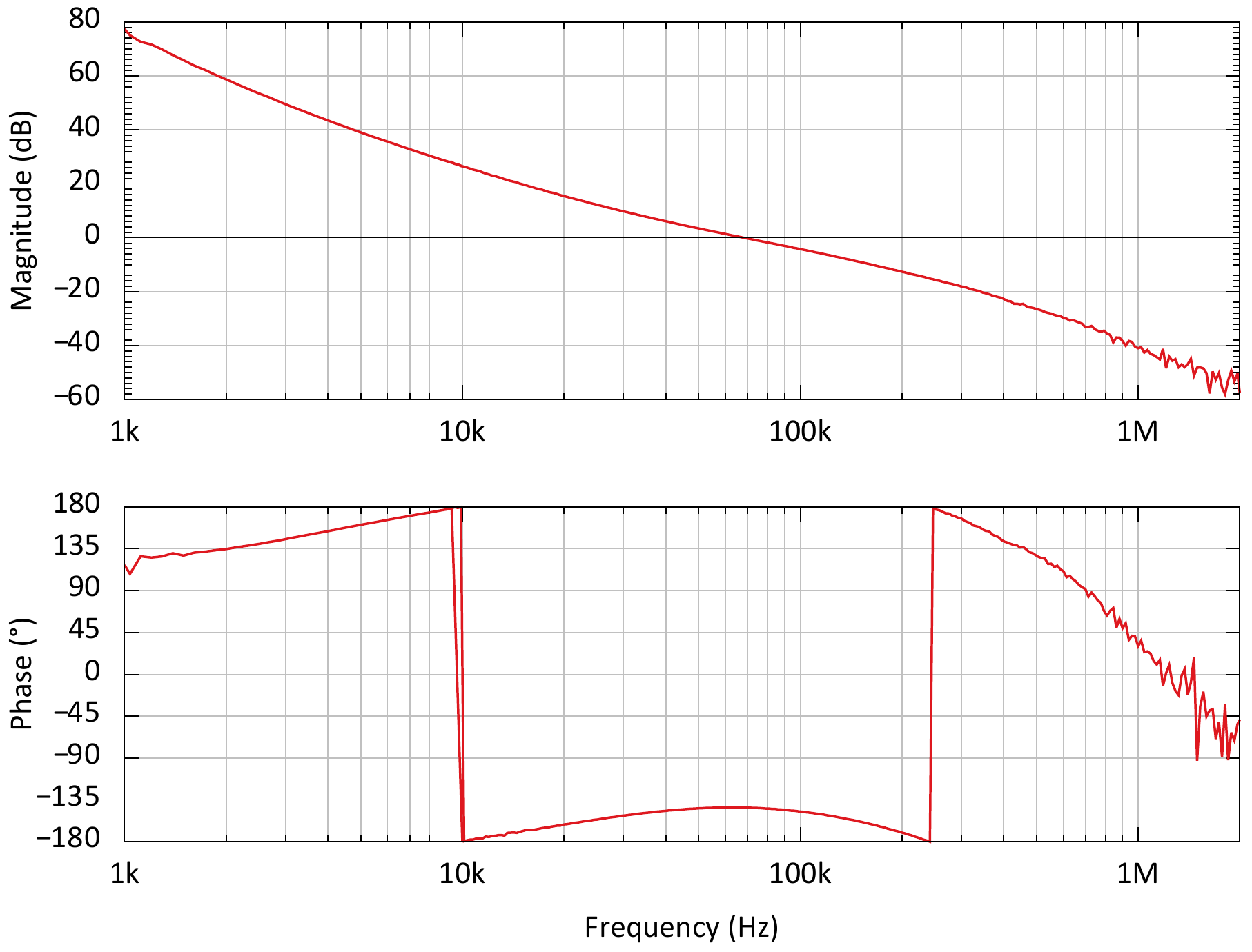}
\caption{{The open} 
 loop transfer function of the power stabilization is designed to operate with a bandwidth of \SI{70}{\kilo\hertz} for good noise performance in the gravitational wave frequency band and for high~robustness.}
\label{fig:openloop_rpn}
\end{figure}

The power stabilization of the laser system used the pump current of the BOA as an actuator. As~an in-loop sensor, a~low noise photodiode was implemented.
The open loop transfer function is designed for a low noise performance in the full gravitational wave frequency band (see Figure~\ref{fig:openloop_rpn}). A~control bandwidth of \SI{70}{\kilo\hertz} with good phase and gain margin for best noise performance and high robustness was chosen. The~bandwidth was limited by the used analog controller and the chosen photodiode electronics. The~in-loop error point noise of the power stabilization was suppressed for frequencies below \SI{10}{\kilo\hertz} by more than one order of magnitude to a value lower than the shot~noise.

\vspace{-3pt}
\begin{figure}[H]
\includegraphics[width=0.95\linewidth]{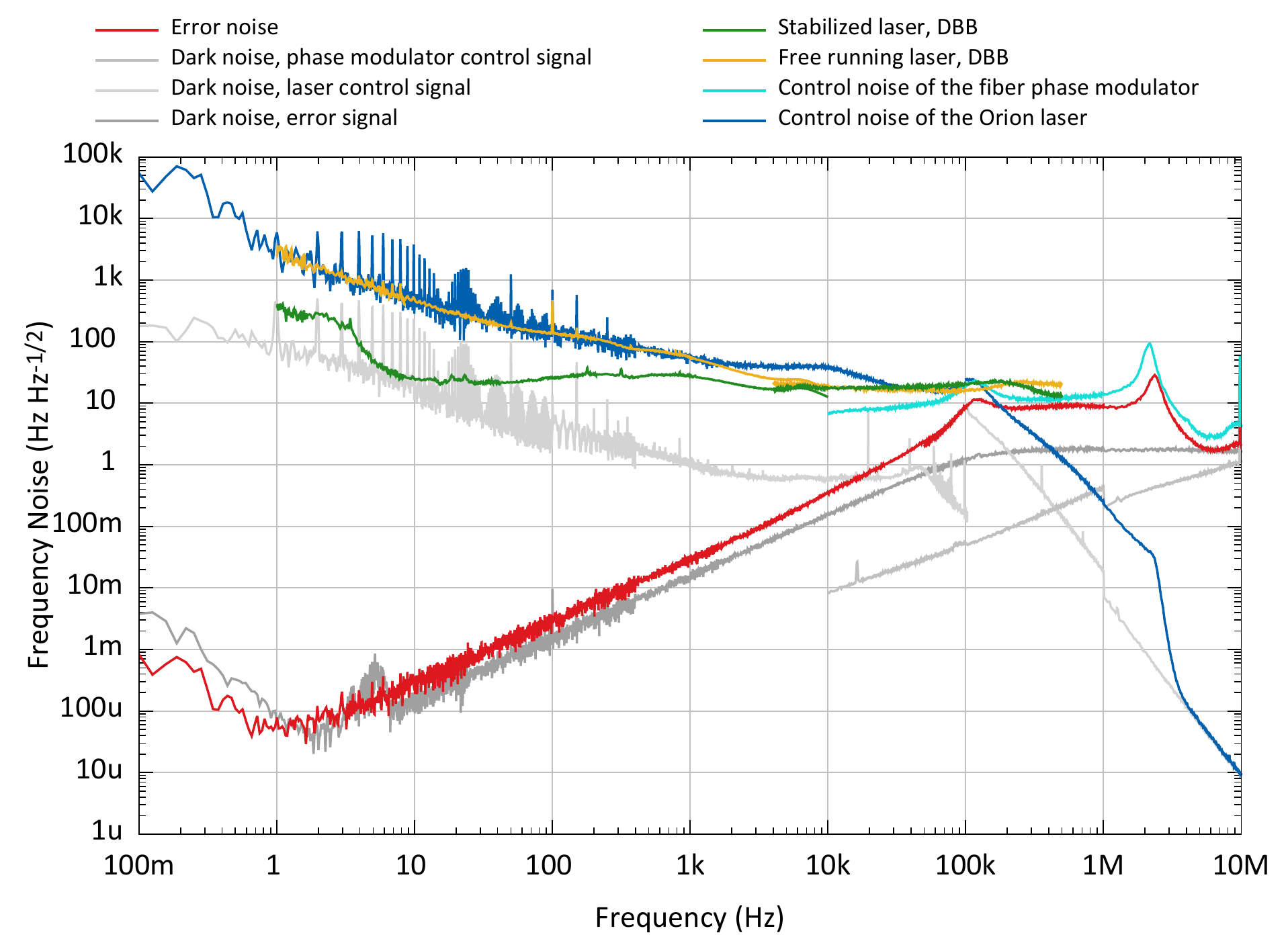}
\caption{The calibrated control signal noise (blue) and the calibrated error signal noise (red) are plotted for the frequency~stabilization. }
\label{fig:stab_frq}
\end{figure}

\vspace{-9pt}

\begin{figure}[H]
\includegraphics[width=0.95\linewidth]{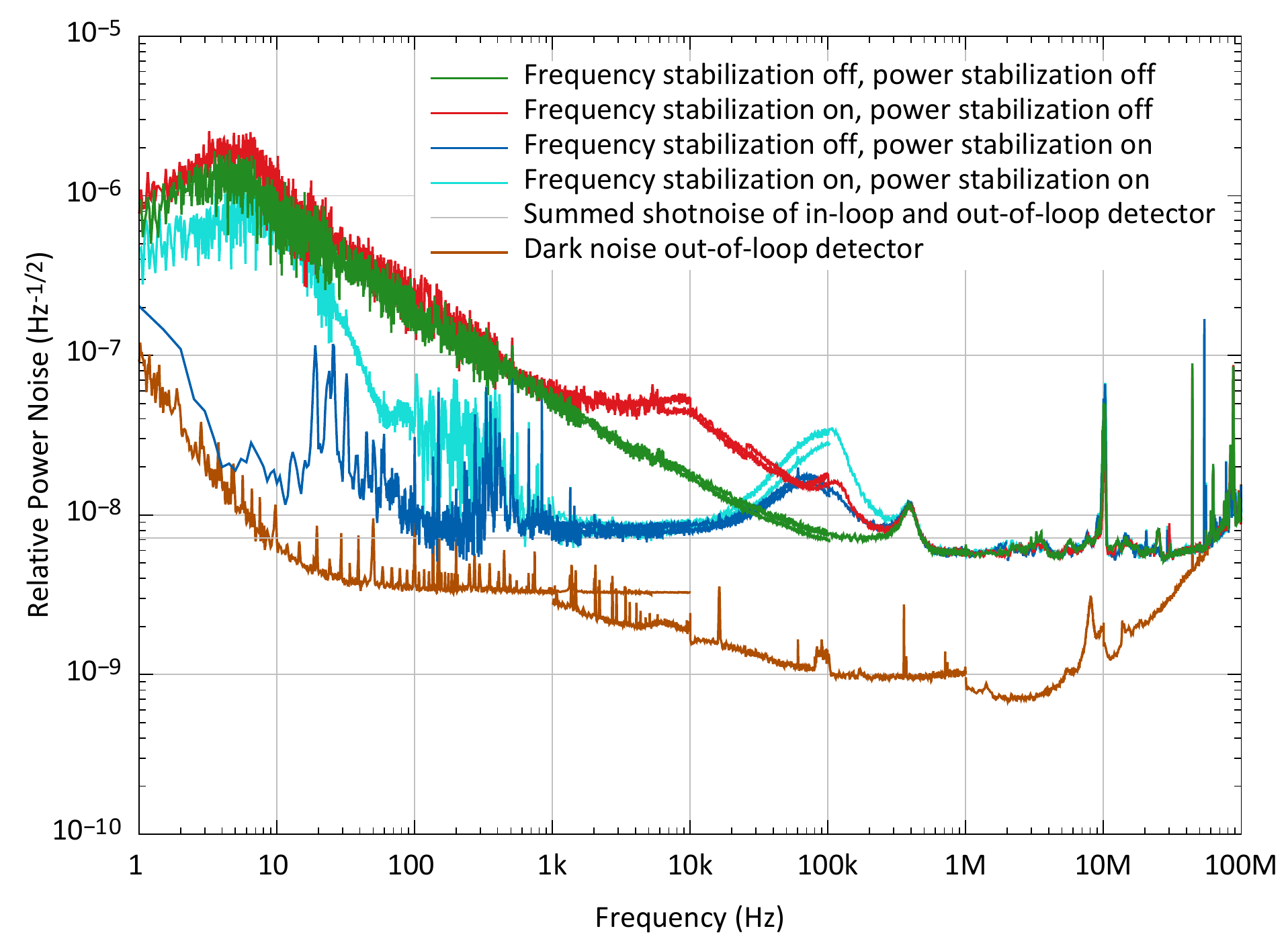}
\caption{{The RPN} 
 as an amplitude spectral density was measured for several combinations of on/off states of the frequency and power stabilization control loops. A~cross-coupling between frequency stabilization to power noise is~visible.}
\label{fig:stab_rpn}
\end{figure}

The power noise measurement photodiode inside the DBB was used as an out-of-loop sensor with \SI{16.3}{\milli\watt} detected power.  The~noise spectra are taken in different operation states of the laser systems, see Figure~\ref{fig:stab_rpn}. At~low frequencies, the~power noise measurements with enabled power stabilization were influenced by non-stationary contributions from scattered light and air turbulences. These noise sources couple as sensor noise into the in-loop and out-of-loop sensors.
In particular, the difference between the measurements with running power stabilization in Figure~\ref{fig:stab_rpn} below \SI{100}{\hertz} is caused by the non-stationary noise contributions, which show up by strong differences of the single traces in the averaged measurements. There was no systematic difference observed between the noise at these low frequencies with and without enabled frequency stabilization while operating the power~stabilization.

\vspace{-6pt}

Between \SI{1}{\kilo\hertz} and \SI{30}{\kilo\hertz}, the~uncorrelated sum of the in-loop detector shot noise and the out-of-loop detector shot noise was the limit of the stabilization. At~\SI{100}{\kilo\hertz}, the servo bump of the stabilization is present, and at \SI{300}{\kilo\hertz} a noise bump of the current driver of the amplifier is visible. At~higher frequencies, the~measurement was limited by the shot noise of the out-of-loop detector and noise peaks from the timing signal of the data acquisition system and by peaks at the modulation frequency.
The noise floor above \SI{1}{\kilo\hertz} was confirmed with measurements using an independent photodiode outside of the DBB as another out-of-loop~detector. 

With an enabled frequency stabilization, the free-running RPN increases between \SI{1}{\kilo\hertz} and \SI{200}{\kilo\hertz}  (see  Figure~\ref{fig:stab_rpn}). This is due to the residual amplitude modulation (RAM) of the fiber phase modulator. The~conversion from phase actuation to laser frequency $\Delta f_L = \Delta\phi\cdot f $ leads to an increased phase actuation $\Delta\phi$ necessary for lower Fourier frequencies $f$. This resulted in the higher RAM in the range of the cross-over frequency. At~these frequencies, higher modulation amplitudes are expected for the laser and the phase actuators, which implies an increased RAM injection into the power stabilization~loop.

To lower the RAM, a~bias voltage of \SI{-5.5}{\volt} was applied to the fiber phase modulator.
A fiber polarization beam splitter behind the modulator is used to clean the polarization, which simultaneously converts polarization modulation into RAM. Only negligible polarization fluctuations remain on the beam after this element such that no significant RAM generation is expected at subsequent polarizing components.
The RAM of the seed beam was further lowered in the amplifier by \SI{20}{\decibel} at low frequencies according to the amplifier transfer function (see Figure~\ref{fig:tfs_power_crossLowAmp}).

When the power stabilization was enabled in addition to the frequency stabilization, the~excess power noise between \SI{2}{\kilo\hertz} and \SI{20}{\kilo\hertz} due to RAM could be suppressed. The~servo bump around \SI{100}{\kilo\hertz} was, however, increased.
With such a high loop gain, the~out-of-loop performance of the power stabilization in the gravitational wave frequency range is only limited by the in-loop sensor noise, which can be improved by better shielding and by detecting more~power.

\section{Conclusions and~Outlook}

Currently, designed future ground-based GWDs operating at a wavelength of \SI{1550}{\nm} demand single mode, single frequency laser systems~\cite{ETST2020} with extremely low noise performance. 
In this paper, we presented detailed measurements of the free-running noise and actuator transfer functions of a variety of seed lasers, pre-amplifiers, and power amplifiers as optional subcomponents for such laser~systems.

Our measurements give an independent characterization and an extension to vendor data sheets to support the trade-off between different lasers for future GWDs or other applications with low noise requirements.
The frequency noise of the tested amplifier systems is defined by the seed laser's frequency noise. Similarly, the~frequency modulation transfer function from the seed to the amplifier output depends primarily on the seed laser's frequency actuators, up~to a delay due to additional propagation time through the amplifiers. 
The power noise of all analyzed MOPA combinations at frequencies below \SI{10}{\kilo\hertz} is dominated by the high power stages, while at higher frequencies, the~power noise of one of the seed lasers starts to dominate (see Figure~\ref{fig:rpnHighAmp}).

In Section~\ref{sec:laserStab}, the performance of the Orion BOA combination in a stabilization environment is discussed. We demonstrate the high bandwidth and high noise suppression of a frequency stabilization with respect to a high finesse optical reference cavity. 
Simultaneously, we operate a power stabilization and show a high loop gain in the detection band of ground-based GWDs. The~performance was limited by non-stationary below \SI{100}{\hertz}.

The final laser system can benefit from an optical mode cleaning resonator by stray light shielding and passive noise filtering above its pole frequency. This also would help to reduce the non-stationary RPN observed below \SI{100}{\hertz}. Such a spatial mode filtering and pointing suppression of an optical mode cleaning resonator might become even more relevant for the further amplified beam of a final laser~system.

Due to the unavailability of an appropriate power actuator for the commercial high power amplifiers, the~light power in our stabilization experiment was limited to the output power of the pre-amplifiers to \SI{115}{\milli\watt}. 
To use the high power stages in future PSL systems, either appropriate power modulation inputs need to be added to the high power amplifiers, which was demonstrated in~\cite{Thies2019} for solid state amplifiers, or~external modulators must be~implemented. 

To integrate the laser system to the stabilizations of a GWD, use of nested control loops and  multi-sensor feedback control can be applied. This combines the fast internal feedback stabilizations of the PSL with the high precision sensors for laser power and laser frequency in the main vacuum system of a GWD~\cite{Kwee2012}.

To reach higher power levels at \SI{1550}{\nm} wavelength, as~demanded  for some future GWDs, additional fiber amplifiers can be used, as~demonstrated in~\cite{Varona2017}.

\vspace{6pt} 



\authorcontributions{\textls[-10]{{Conceptualization, F.M., B.W.; methodology,  F.M., B.W.; investigation F.M.; writing---original draft preparation,  F.M.; writing---review and editing, B.W. All authors have read and agreed to the published version of the manuscript.}}
}

\funding{Funded by the Deutsche Forschungsgemeinschaft (DFG, German Research Foundation) under Germany’s Excellence Strategy{---}
EXC-2123 QuantumFrontiers{---}390837967.}


%

 \dataavailability{Not applicable.}

\conflictsofinterest{The authors declare no conflict  of~interest.}

\begin{adjustwidth}{-\extralength}{0cm}

\reftitle{References}




\end{adjustwidth}
\end{document}